\documentclass[twocolumn]{raa}
 
\usepackage{graphicx,times}             
\usepackage{natbib}
\usepackage{amssymb,amsmath}
\usepackage{mathrsfs}            
\bibpunct{(}{)}{;}{a}{}{,}

\usepackage[pagebackref=true]{hyperref}

\newcommand{\mpc}{{\, {\rm Mpc}}}

\def\apj{ApJ}
\def\apjs{ApJS}

\def\mnras{MNRAS}
\def\aap{A\&A}

\def\nat{Nature}      
\def\apjs{ApJS}
\def\apjl{ApJ Letters}

\begin{document}

\title{Galaxy interactions in filaments and sheets: insights from EAGLE simulations}

   \volnopage{Vol.0 (20xx) No.0, 000--000}     
   \setcounter{page}{1}           

    \author{Apashanka Das
      \inst{1}
   \and Biswajit Pandey
      \inst{2}
      \and Suman Sarkar
      \inst{3,4}
   }
 \institute{Department of Physics, Visva-Bharati University, Santiniketan, 
              Birbhum, 731235, India
             {\it a.das.cosmo@gmail.com}\\
        \and
	     {Department of Physics, Visva-Bharati University, Santiniketan, 
	     Birbhum, 731235, India {\it biswap@visva-bharati.ac.in}}\\
        \and
            {Department of Physics, Indian Institute of Science Education and
              Research Tirupati, Tirupati - 517507, India
        \and
            {Department of Physics, Indian Institute of Technology Kharagpur, Kharagpur, 721302, India}
              {\it suman2reach@gmail.com}}
   }
       
\vs\no
   {\small Received 20xx month day; accepted 20xx month day}

\abstract{We study the colour and star formation rates of paired
  galaxies in filaments and sheets using the EAGLE simulations. We
  find that the major pairs with pair separation $<50$ kpc are bluer
  and more star forming in filamentary environments compared to those
  hosted in sheet-like environments. This trend reverses beyond a pair
  separation of $\sim 50$ kpc. The interacting pairs with larger
  separations ($>50$ kpc) in filaments are on average redder and
  low-star forming compared to those embedded in sheets. The galaxies
  in filaments and sheets may have different stellar mass and cold gas
  mass distributions. Using a KS test, we find that for paired
  galaxies with pair separation $<50$ kpc, there are no significant
  differences in these properties in sheets and filaments. The
  filaments transport gas towards the cluster of galaxies. Some
  earlier studies find preferential alignment of galaxy pairs with
  filament axis. Such alignment of galaxy pairs may lead to different
  gas accretion efficiency in galaxies residing in filaments and
  sheets.  We propose that the enhancement of star formation rate at
  smaller pair separation in filaments is caused by the alignment of
  galaxy pairs. A recent study with the SDSS data \citep{das23}
  reports the same findings. The confirmation of these results by the
  EAGLE simulations suggests that the hydrodynamical simulations are
  powerful theoretical tools for studying the galaxy formation and
  evolution in the cosmic web.  }

\authorrunning{A. Das, B. Pandey \& S. Sarkar}
\titlerunning{Galaxy interactions in the cosmic web}
 
\maketitle

\section{Introduction}           
\label{sect:intro}

The galaxies are the fundamental units of the observed large-scale
structures in the Universe. Understanding their formation and
evolution is one of the primary goals of modern cosmology.  The growth
of the primordial density perturbations via the gravitational
instability eventually leads to the formation of the dark matter
halos. The dark matter halos represent the peaks in the density
field. The halos are surrounded by diffuse neutral Hydrogen
distribution after the recombination. They accrete the gas which
radiate away their kinetic energy and settle down at their
centers. The cooling and condensation of the gas at the centre of
these halos are believed to be the primary mechanism for galaxy
formation \citep{silk77, white78}.

The formation and evolution of galaxies are expected to be influenced
by both the initial conditions at the location of their formation and
their interactions with the surrounding environment.  The galaxies
interact with other galaxies in their neighbourhood. The galaxy-galaxy
interactions are known to amplify the star formation rate in galaxies
\citep{barton00, nikolic04, woods07, ellison10, patton11}. The
environments of galaxies have crucial roles in their evolution. The
colour and star formation rates of galaxies are strongly affected by
the local density of their environment \citep{gomez03, lewis02}. The
galaxies become redder and low-star forming in the higher density
environments \citep{kauffmann04}. The suppression of star formation
can be driven by different physical mechanisms. The ram pressure
stripping is a common phenomena in galaxy clusters \citep{gunn72}. The
galaxies in the high density regions are more likely to encounter
harassment \citep{moore96, moore98}, starvation \citep{larson80,
  somerville99}, strangulation \citep{gunn72, balogh00} and gas
expulsion by supernovae, AGN or stellar winds \citep{cox04, murray05,
  springel05}. The star formation in galaxy may also quench through
several other routes. The mass \citep{birnboim03, dekel06}, morphology
\citep{martig09}, presence of bar \citep{masters10} and high angular
momentum \citep{peng20} can cease the star formation activity in
galaxies.

Many other galaxy properties depend on the environment. The elliptical
galaxies are more commonly observed in dense groups and clusters
\citep{oemler74, dress80, davis76, guzo97, gotto03}. The spiral
galaxies mostly occupy the intermediate and low density regions of the
Universe. These environmental dependencies of clustering are reflected
in different statistical measures such as the correlation function
\citep{zehavi}, genus \citep{park1}, filamentarity \citep{pandey05,
  pandey06}, local dimension \citep{pandey20} and mutual information
\citep{pandey17, bhattacharjee20, sarkar20}. The environment of a
galaxy is generally characterized by the local density. The local
density undoubtedly plays a decisive role in galaxy
evolution. However, it can not completely characterize the environment
of a galaxy. The early generation redshift surveys reveal that
galaxies are part of an all-inclusive network comprising clusters,
filaments and sheets surrounded by vast empty regions
\citep{joeveer78, gregory78, einasto80, zeldovich82}. The galaxies and
their host halos are embedded in different environments of the cosmic
web \citep{bond96}. \citet{pandey08} find that the star forming blue
galaxies trace a more filamentary distribution compared to the red
galaxies. More than $80\%$ of the baryonic budget in the Universe is
accounted by low density gas (WHIM) in filaments \citep{tuominen21,
  galarraga21}. Consequently, the gas accretion efficiency of the dark
matter halos in different environments may differ in a significant
manner. Thus, the cosmic web can have significant impact on the galaxy
properties and their evolution. The galaxies that are located in
different parts of the cosmic web can experience different physical
conditions, such as different densities of gas, different levels of
tidal forces, different frequency of interactions and mergers.

The interactions between galaxies with comparable masses are known as
the major interactions. Such interactions trigger new star formation
in galaxies. The interacting pairs can be hosted in different
morphological environments of the cosmic web. The galaxy pairs are
more frequently observed in the denser regions. The filaments and
sheets, being the denser parts of the cosmic web, can host a
significant number of major galaxy pairs. In a recent work,
\citet{das23} analyze the SDSS data to compare the star formation rate
and colour of major pairs hosted in filaments and sheets. They find
that the major galaxy pairs with separation $<50$ kpc are relatively
high star forming and bluer when hosted in filaments. Contrarily, the
major pairs at separations larger than $50$ kpc show a significantly
higher SFR and bluer colour in sheet-like environments.  This
behaviour may be related to the preferential alignment of galaxy pairs
with the filament axis reported in a number of recent works
\citep{tempel15, mesa18}. The star formation in a galaxy is primarily
regulated by its available gas mass. The inflows and outflows
\citep{dekel09, dave12} can significantly modulate the gas mass in a
galaxy. The transient events like interactions and mergers can drive
the galaxies out of equilibrium. The alignment of galaxy pairs with
filament spines may lead to anisotropic accretion and higher gas
accretion efficiency in these galaxies. In this work, we intend to
verify these findings using hydrodynamical simulations.

The EAGLE simulation \citep{eagle16} is a hydrodynamical simulation
that studies the galaxy formation and evolution in a cosmological
volume. It describes the formation of galaxies by gas falling into the
dark matter halos and their subsequent cooling and condensation. It
would be interesting to study the colour and star formation rate in
the major pairs in filaments and sheets using EAGLE simulations. In
observations, the galaxy pairs are usually identified by applying
simultaneous cuts on the projected separation and the velocity
difference of the galaxies in the rest frame. However, all these pairs
may not be undergoing interactions. Some of the pairs identified in
observations may not be close in three dimensions due to the chance
superposition in the high-density regions like groups and clusters
\citep{alonso04}. Also, we can not construct a mock catalogue for the
observational sample of galaxy pairs used in \citet{das23} due to the
smaller volume of the EAGLE simulations. So we decided to use the
real-space positions of galaxies available in simulation to identify
the major pairs. This would avoid any errors in identification of
galaxy pairs due to the projection effects. We identify the geometric
environments of galaxy pairs using the local dimension
\citep{sarkar09}. Our primary aim of this work is to study the
interaction induced star formation in filaments and sheets using EAGLE
simulations. This would help us to asses the roles of the filaments
and sheets in galaxy evolution.

We organise the structure of the paper as follows: we describe the
data and the method of analysis in Section 2 and present the results
and conclusions in Section 3.

\section{Data and Method of Analysis}           

\subsection{EAGLE simulation data}
\label{sec:data}

The EAGLE simulation \citep{eagle16} is a set of cosmological
hydrodynamical simulation in periodic, cubic comoving volumes ranging
from side of length 25 to 100 Mpc. It tracks the evolution of both
baryons and dark matter in the Universe from a redshift of 127 to
0. The simulation adopts a flat $\Lambda CDM$ cosmology with
$\Omega_{\Lambda}=0.693$, $\Omega_m=0.307$, $\Omega_b=0.04825$ and $H_0=67.77$ km/s/Mpc \citep{planck14}.

We download the various properties of galaxies from the publicly
available EAGLE run simulation. We extract the information of position
of the centre of mass of galaxies in three dimensions within a
comoving cubic volume of size $100$ Mpc$^3$ from
$Ref-L0100N1504\_Subhalo$ table. We consider the last snapshot of the
simulation having $Snapnum=28$ which corresponds to redshift $z=0$. We
select only those galaxies which are flagged as $Spurious=0$. This
ensures that we select only the genuine simulated galaxies by
discarding all the unusual objects with anomalous stellar mass,
metallicity or black hole mass. We also download the star formation
rate, stellar mass and cold gas mass of simulated galaxies using
$Ref-L0100N1504\_Aperture$ table. These are estimated within a
spherical $3D$ aperture of radius 30 kpc centered at the location of
the minimum gravitational potential of a galaxy. Use of this criteria
gives well suited stellar mass and star formation estimates as
compared to observational results and is also recommended for use by
the EAGLE simulation team \citep{eagle16}.  We also consider only
those galaxies with a stellar mass $>0$. Combining the two tables with
$GalaxyID$ , we obtain all of the above mentioned information for
325358 galaxies. We also extract the rest frame broadband magnitudes
of galaxies estimated in $u$ and $r$ band filters \citep{doi10} from
$Ref-L0100N1504\_Magnitude$ table, where $u$ and $r$ respectively
denote Ultraviolet and Red filter bands of Sloan Digital Sky Survey
(SDSS). We combine this table with $Ref-L0100N1504\_Subhalo$ and
$Ref-L0100N1504\_Aperture$ table using $GalaxyID$ to get all the
required information. The magnitude of galaxies in different SDSS
filters are also computed in 30 kpc spherical apertures
\citep{trayford15}. Finally, we have all these information for 29754
galaxies. For the rest of the analysis, we refer to $u-r$ colour of
galaxies as the difference of its rest frame non-dust attenuated
absolute magnitudes in $u$ and $r$ band respectively. Only the
magnitudes of the galaxies with stellar mass $log(M_{stellar}/M_{sun})
> 8.5$ are provided in $Ref-L0100N1504\_Subhalo$ table.  However, here
we use the stellar mass estimates of galaxies from
$Ref-L0100N1504\_Aperture$ table where the minimum stellar mass of a
galaxy is $log(M_{stellar}/M_{sun}) \sim 8.2$.  Observations show that
the galaxies with stellar mass $M_{stellar}<3 \times 10^{10}\,M_{sun}$
are actively star forming and the galaxies having stellar masses above
this critical value are generally quiescent systems
\citep{kauffmann03}. For the present analysis, we consider only those
galaxies which have their stellar mass in between $8.5 \leq
log(M_{stellar}/M_{sun}) \leq 10.5$. Our mass limited sample contains
a total 21305 galaxies.

We identify the nearest neighbour in three dimensions for each galaxy
in our mass limited sample. We denote the distance to the nearest
neighbour for each galaxy by $r$. Here distance refers to the 3D
physical separation between the centre of mass of the
galaxies. Initially, we label each galaxy and its nearest neighbour in
our sample as a possible pair. We then select only those pairs for
which $r \leq 200$ kpc. We also apply a cut on their stellar mass
ratio $1 \leq \frac{M_{1}}{M_{2}} < 3$ to include only the major pairs
in our analysis. This provides us with a total 2264 major pairs.  The
smallest pair separation in our sample is $\sim 6$ kpc.

 We determine the morphological environment of the galaxies in the
 EAGLE simulation by estimating their local dimension
 (\autoref{subsec:ldim}). We use $GalaxyID$ to cross match these
 galaxies with our pair sample. The cross-matching yields a total 2537
 galaxies in major pairs. We find that 373 and 276 out of these
 galaxies are residing in filaments and sheets respectively. It may be
 noted that we can not determine the local dimension of all the
 galaxies in the simulation.

\begin{table}
\centering
\begin{tabular}{|c|c|}
\hline
Morphological environment & Local dimension\\
\hline
$D1$ & $0.75 \leq D < 1.25$ \\
$D2$ & $1.75 \leq D < 2.25$  \\
$D3$ & $D \geq 2.75$ \\
$D1.5$ & $1.25 \leq D < 1.75$ \\
$D2.5$ & $2.25 \leq D < 2.75$  \\
\hline 

\end{tabular}
\caption{This table shows the definition of different geometric
  environments based on the local dimension (D) of the galaxies.}
\label{tabld}
\end{table}

\subsection{Geometry of the local environment}
\label{subsec:ldim}
We characterize the different geometric environments of the cosmic web
using the local dimension \citep{sarkar09}. The local dimension is a
simple measure based on the galaxy counts within spheres of different
radii centered on a galaxy. The galaxy counts within a sphere of
radius $R$ centered on a galaxy can be written as,
\begin{equation}
N(< R) = A\,R^D
\label{ld}
\end{equation}
where $D$ is the local dimension and $A$ is an arbitrary constant. The
number counts $N(< R)$ would scale differently with the radius $R$
depending on the local geometry of the embedding environment. The
radius of the sphere around each galaxy is varied between $R_1 \mpc
\leq R \leq R_2 \mpc$ and the galaxy counts are measured for each
radius. Only the galaxies that have at least 10 neighbouring galaxies
within this range are considered for this analysis. We fit the
observed number counts $N(<R)$ to \autoref{ld} using a least-square
fitting. The goodness of each fit is determined by estimating the
$\chi^2$ per degree of freedom. We only retain the fits having
$\frac{\chi^2}{\nu} \leq 0.5$ \citep{sarkar19} and discard the rest
from our analysis. We choose $R_1=2 \mpc$ and $R_2=10 \mpc$ for the
analysis presented here.

The local dimension $D$ describes the morphology of the embedding
environment. Ideally, a filamentary environment should have $D=1$ and
sheet-like environment should have $D=2$. A homogeneous distribution
in three-dimension is represented by $D=3$. However, the filaments,
sheets, clusters and voids are not idealized structures and they can
have a wide variety of shapes and sizes. Each geometric environment is
assigned a range of local dimension \autoref{tabld}. A nearly straight
filament is represented by $D1$-type environment. Similarly, the
$D2$-type environment represents a two-dimensional sheet-like
structure. A $D3$-type environment is embedded in a three dimensional
distribution of homogeneous nature. The galaxies can also reside near
the junction of different types of morphological environments. $D1.5$
and $D2.5$ can be treated as intermediate environments.

\begin{figure*}
\resizebox{14.4cm}{6cm}{\rotatebox{0}{\includegraphics{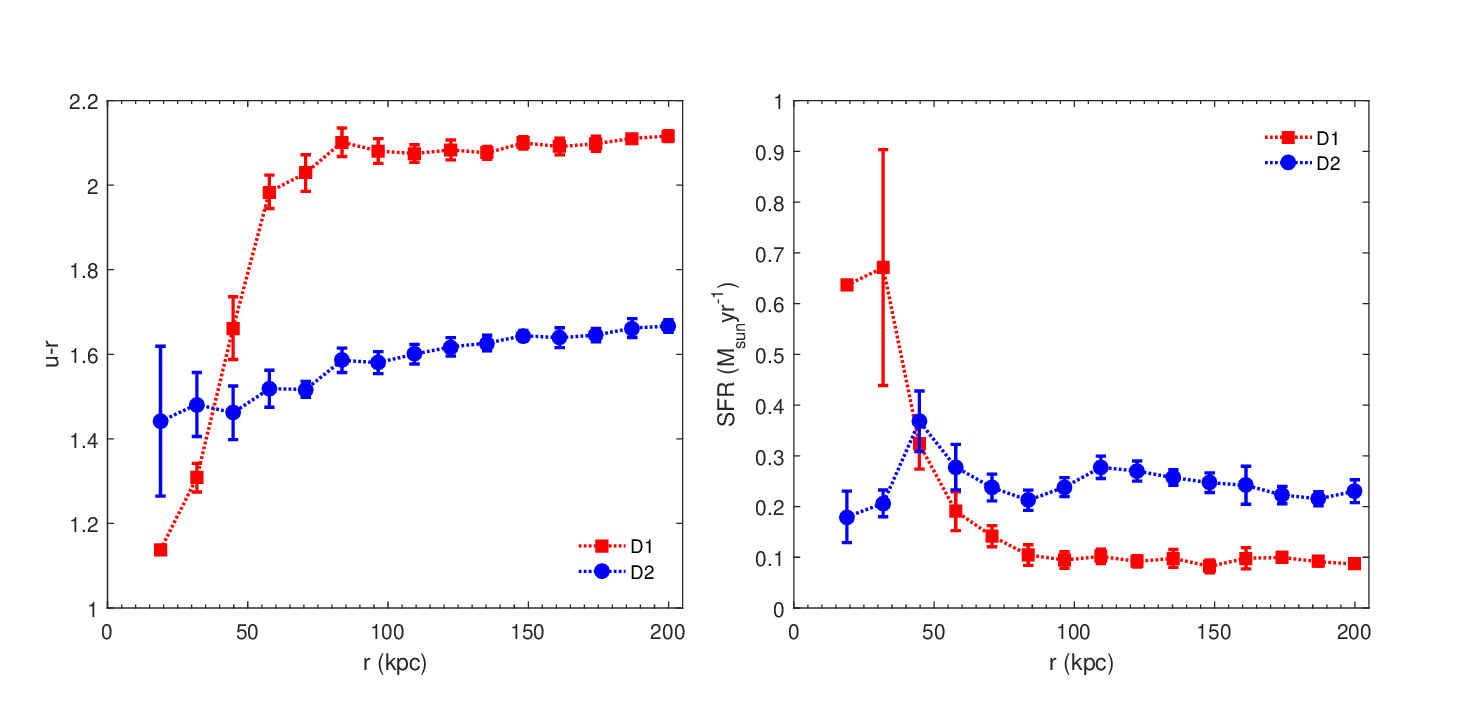}}}
\caption{The left panel of this figure shows the cumulative mean $u-r$
  colour as a function of pair separation for major pairs residing in
  $D1$ and $D2$ type environments. The right panel shows the
  cumulative mean SFR for the same pairs in two different
  environments.  We use 10 Jackknife samples to estimate the $1\sigma$
  error bars shown at each data point.}
\label{Fig1}
\vspace{0.2cm}
\end{figure*}

\begin{figure*}
\resizebox{14.4cm}{6cm}{\rotatebox{0}{\includegraphics{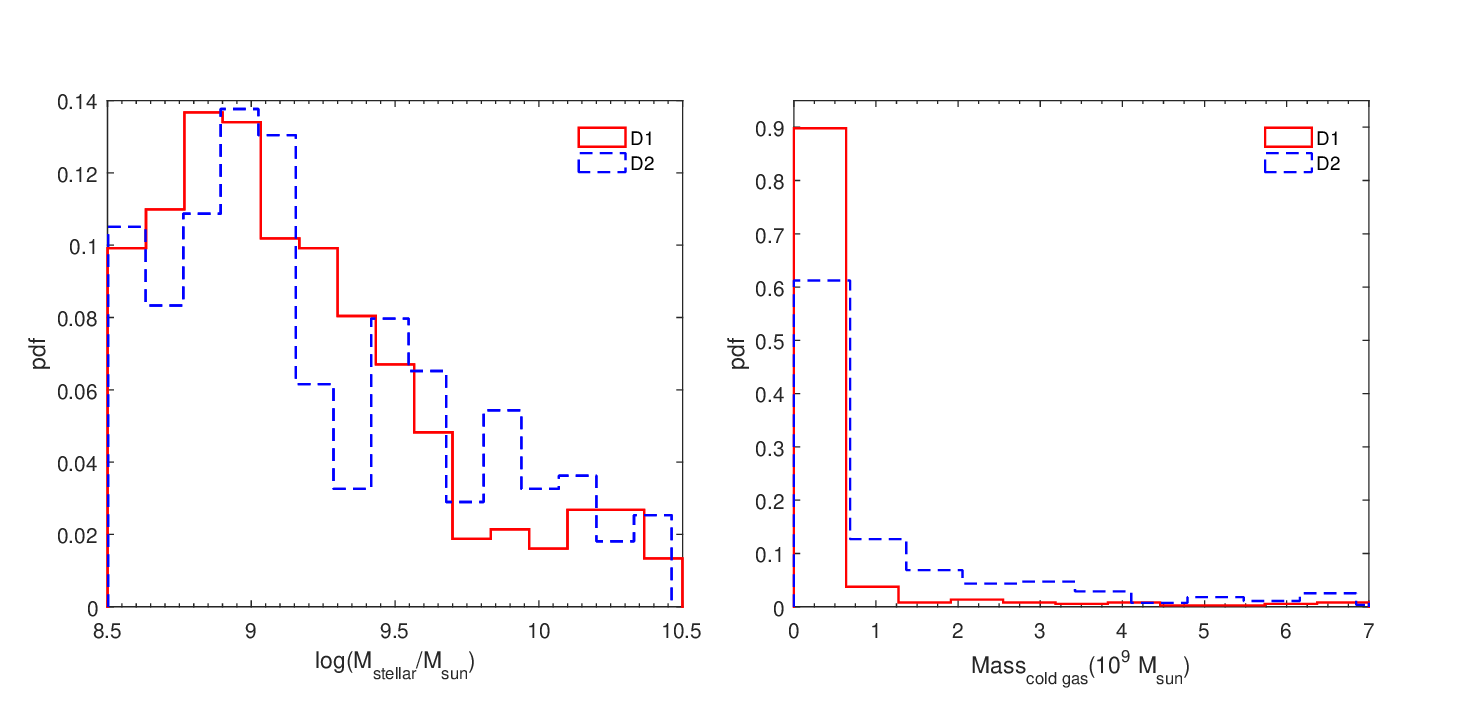}}}
\caption{The left panel shows the probability distribution function of
  $log(M_{stellar}/M_{sun})$ for major pairs in $D1$ and $D2$ type
  environments. The right panel shows the probability distribution
  function of $Mass_{cold gas}$ for the same pairs.}
\label{Fig2}
\end{figure*}



\begin{figure*}
\resizebox{8cm}{6cm}{\rotatebox{0}{\includegraphics{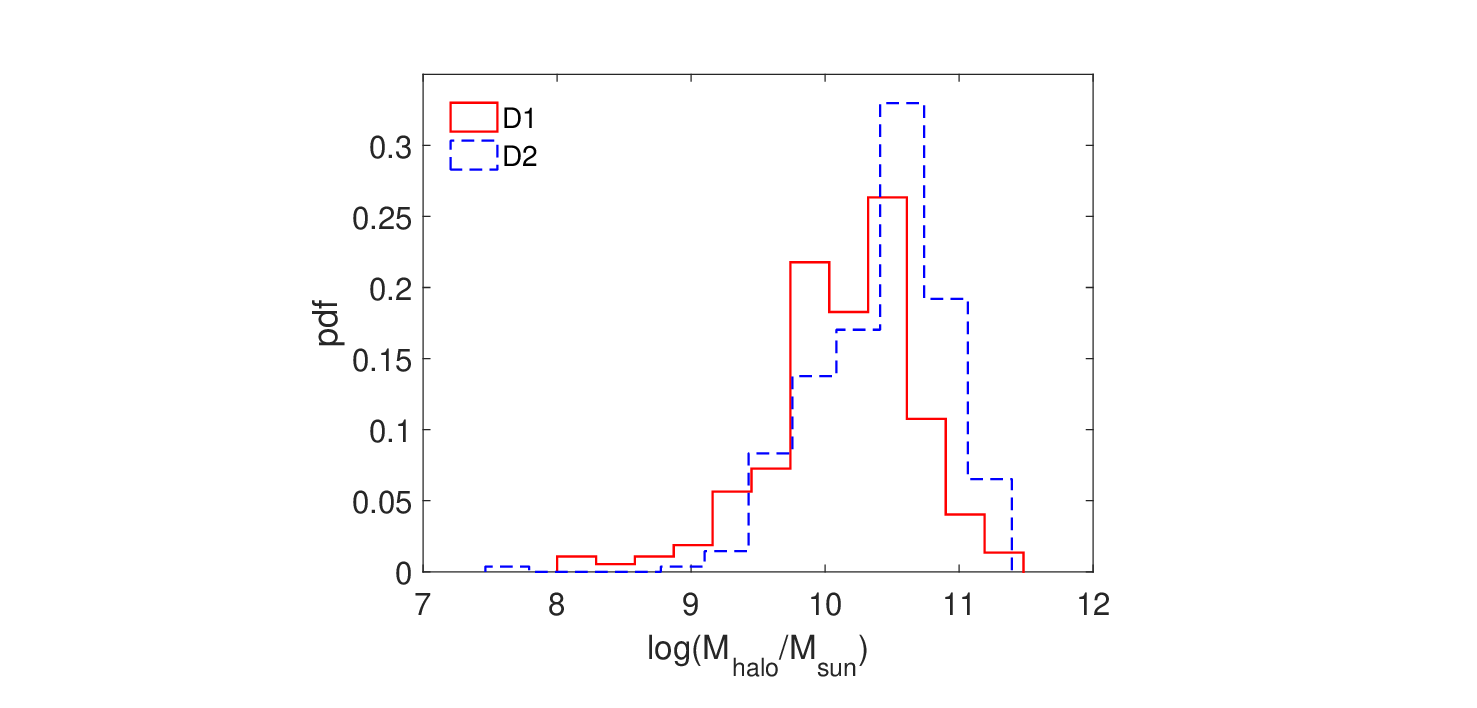}}}
\caption{This figure shows the distributions of the halo mass of
  paired galaxies in sheets and filaments.}
\label{Fig3}
\vspace{0.2cm}
\end{figure*}


\begin{figure*}
\resizebox{14.4cm}{6cm}{\rotatebox{0}{\includegraphics{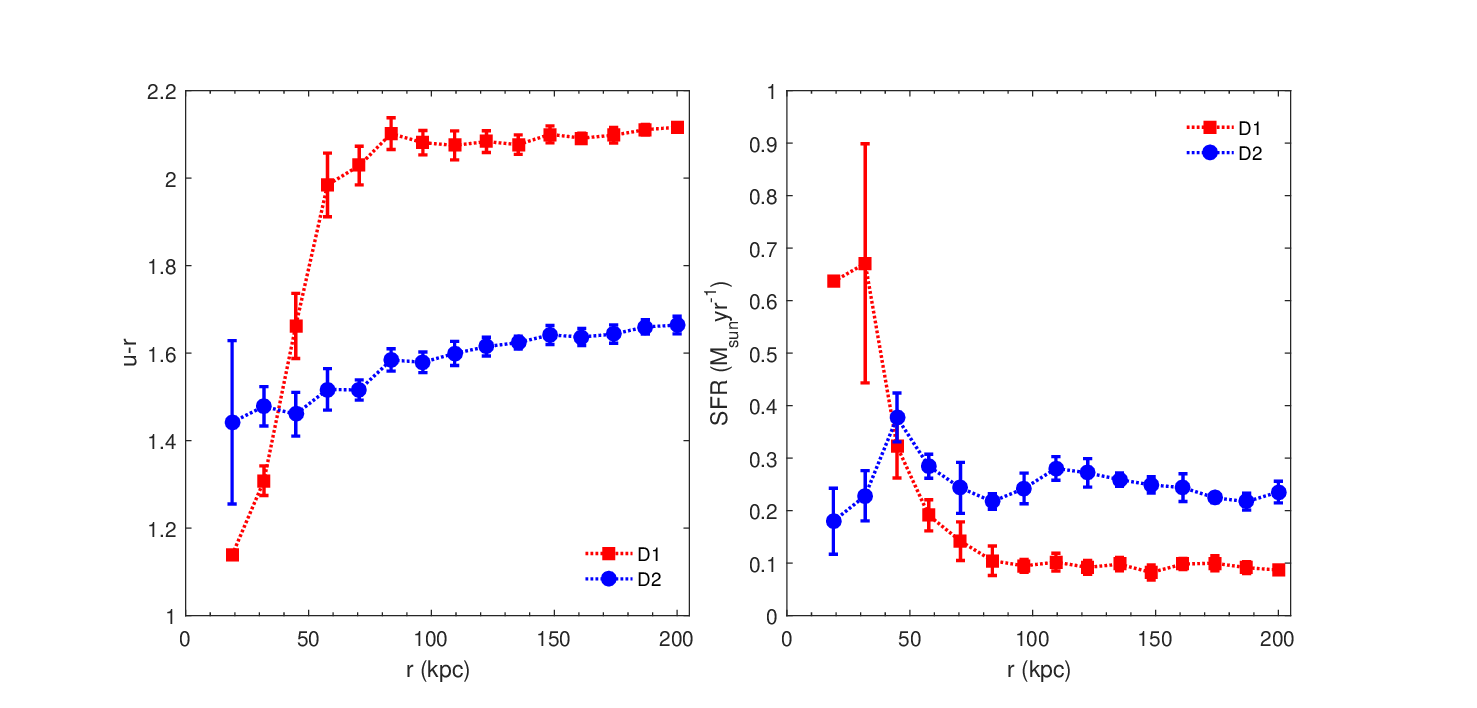}}}
\caption{Same as \autoref{Fig1} but when the galaxy positions are
  defined by the location of the minimum of the gravitational
  potential instead of the centre of mass.}
\label{Fig4}
\vspace{0.2cm}
\end{figure*}

\begin{figure*}
\resizebox{14.4cm}{6cm}{\rotatebox{0}{\includegraphics{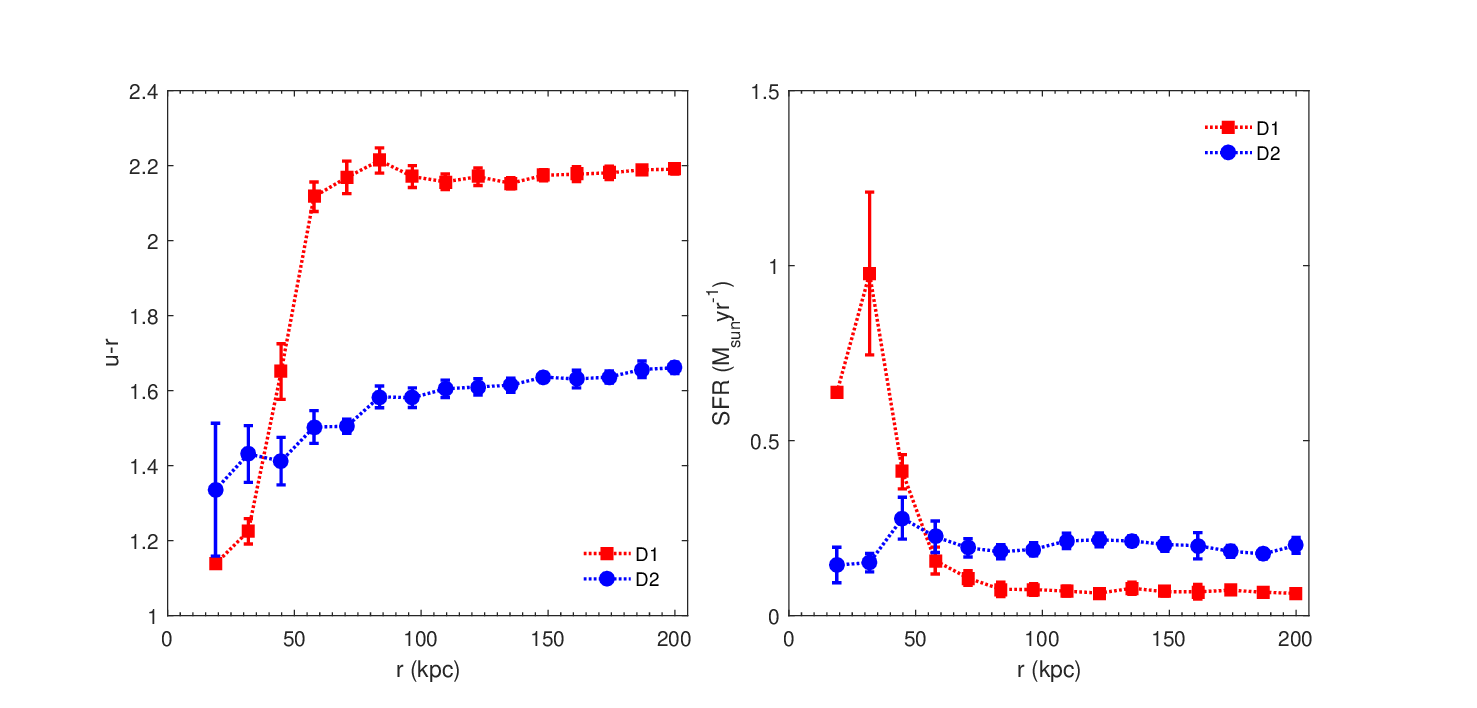}}}
\caption{Same as \autoref{Fig1} but after discarding the major pairs
  for which the $5^{th}$ nearest neighbour lies within a distance of
  $500$ kpc to $1$ Mpc.}
\label{Fig5}
\vspace{0.2cm}
\end{figure*}

\section{Results and Conclusions}

The cumulative mean of the $u-r$ colour for the major galaxy pairs as
a function of the pairs separation is shown in the left panel of
\autoref{Fig1}. We compare the results for the major pairs in
filaments and sheets in the same panel.

This shows that the major pairs with pair separation $r<50$ kpc are on
average bluer in filamentary environment compared to those residing in
sheet-like environment. However, this trend only persists up to a pair
separation of $\sim 50$ kpc. A cross over of the two curves
corresponding to $D1$ and $D2$ type environments is observed at $r
\sim 50$ kpc. The major pairs with pair separation $r>50$ kpc are
significantly redder in filaments compared to those located in
sheets. We also analyze the SFR in major pairs residing in filaments
and sheets and show the results in the right panel of
\autoref{Fig1}. We find that the major pairs at closer pair separation
($<50$ kpc) in filaments are comparatively more star forming than
those located in sheets. We see an exactly opposite trend for the
major pairs with larger pair separation ($>50$ kpc). The colour of the
galaxies are strongly correlated with their SFR \citep{baldry04} and
the results shown in the two panels of \autoref{Fig1} are consistent
with each other. It is also interesting that the crossover is observed
at nearly the same pair separation ($\sim 50$ kpc) for both colour and
SFR. We estimate the 1-$\sigma$ error bars at each pair separation
using $10$ Jackknife samples drawn from the original datasets.

\begin{table*}
\centering
\begin{tabular}{|c|c|c|c|c|c|c|c|}
\hline 
& \multicolumn{2}{c|}{$D_{KS}$} & \multicolumn{5}{c|}{$D_{KS}(\alpha)$}\\
\cline{2-8}
Major pairs & $log(M_{stellar}/M_{sun})$ & $Mass_{cold gas} (10^9\,M_{sun})$ & 99\% & 90\% & 80\% & 70\% & 60\%\\
\hline
All & 0.1037 & 0.5105 & 0.1292 & 0.0972 & 0.0852 & 0.0773 & 0.0712 \\
$ r < 50$ kpc & 0.1032 & 0.3512 & 0.3754 & 0.2826 & 0.2478 & 0.2249 & 0.2072\\
$ r \ge 50$ kpc & 0.1055 & 0.5104 & 0.1388 & 0.1043 & 0.0915 & 0.0830 & 0.0765\\
\hline
\end{tabular}
\caption{This table shows the summary of the Kolmogorov-Smirnov (KS)
  tests carried out for a comparison of $log(M_{stellar}/M_{sun})$ and
  $Mass_{cold gas}$ of major pairs in filaments and sheets. Separate
  comparisons are also carried out for major pairs having $r < 50$ kpc
  and $r \ge 50$ kpc. The table lists the KS statistic $D_{KS}$ along
  with the critical values $D_{KS}(\alpha)$ beyond which the null
  hypothesis can be rejected at a given confidence level.}
\label{tab:ks}
\end{table*}

\begin{table*}
\centering
\begin{tabular}{|c|c|c|c|c|c|c|c|}
\hline 

& \multicolumn{1}{|c|}{$D_{KS}$} & \multicolumn{5}{c|}{$D_{KS}(\alpha)$}\\
\cline{2-7}
Major pairs & $log(M_{halo}/{M_{sun}})$ & 99\% & 90\% & 80\% & 70\% & 60\%\\
\cline{2-7}
  & 0.6532 & 0.1292 & 0.0972 & 0.0852 & 0.0773 & 0.0712 \\
\hline
\end{tabular}
\caption{This table shows Kolmogorov-Smirnov statistic $D_{KS}$ for
  comparison of $log(M_{halo}/M_{sun})$ of major pairs residing in
  $D1$ and $D2$ type environments. It also shows the critical values
  $D_{KS}(\alpha)$ above which the null hypothesis can be rejected at
  different confidence level.}
\label{tab:kshalo}
\end{table*}

The stellar mass \citep{birnboim03, dekel06, bamford09} and the
available cold gas mass content \citep{saintonge12, violino18,
  thorp22} also play a very important role in deciding the star
formation rate in galaxies.  We test if the differences occurring in
$u-r$ colour and SFR of galaxies in major pairs residing in $D1$ and
$D2$ type environments arise due to the differences in their stellar
mass and cold gas content. We use a KS test to compare the
distributions of stellar mass and cold gas mass of major paired
galaxies in $D1$ and $D2$ type environments. The probability
distribution functions of the two properties in $D1$ and $D2$ type
environments are shown in the two panels of \autoref{Fig2}. We first
carry out the test for the major pairs with all possible pair
separations. We then conduct separate tests for the major pairs with
pair separation $>50$ kpc and $<50$ kpc. The results for the KS test
are tabulated in \autoref{tab:ks}.  We note that null hypothesis for
all the major pairs can be rejected at $90\%$ and $99\%$ confidence
levels for stellar mass and cold gas mass respectively. This implies
that the stellar mass distribution of galaxies in major pairs residing
in $D1$ and $D2$ type environment are likely to be drawn from the same
parent population. However, the galaxies in major pairs from filaments
and sheets have a significantly different cold gas mass
distribution. We also arrive at the same conclusions for the major
pairs with $r>50$ kpc. Interestingly, the results for
the major pairs with $r< 50$ kpc suggest that the null hypothesis for
stellar mass can be rejected at a very low confidence level ($<
60\%$), whereas for cold gas mass, it can be rejected at $\leq 90\%$
confidence level. Thus, stellar mass of major pair galaxies with
$r<50$ kpc in $D1$ and $D2$ type environment are highly likely to be
drawn from the same underlying population. This clearly shows that
stellar mass and available cold gas mass of the paired galaxies are
not responsible for the differences observed in their $u-r$ colour and
SFR in $D1$ and $D2$ type environments at smaller pair separations
($r< 50$ kpc).

Each galaxy is believed to have formed within a dark matter halo. The
properties of the galaxy are expected to be intimately connected to
the mass of the dark matter halo. In fact, the mass of the dark matter
halo is believed to be the most important parameter that determines
the properties of a galaxy \citep{cooray}. The amount of substructures
in the dark matter halos increases with the increasing halo mass
\citep{gao04, pandey13}. There are observational evidences in favour
of the correlations between substructure and star formation fraction
in galaxy clusters \citep{bravo09, cohen14}. Substructures can also
influence the stellar population in the galaxy \citep{helmi20}. We
show the distributions of the halo masses in the paired galaxies in
filaments and sheets in \autoref{Fig3}. The halo masses are obtained
within the same aperture as the galaxies.  We perform a KS test to
find that the halo mass distribution of the paired galaxies in sheets
and filaments are significantly different (\autoref{tab:kshalo}). We
find that the halo mass of the paired galaxies in sheets are
relatively more massive than those residing in the filaments. The
effects of the halo mass may also come from the virial shock heating
of the halo gas that becomes important at masses greater than
$10^{12}\,M_{sun}$ \citep{birnboim03}. Such heating can suppress the
supply of cold gas by preventing cold streams from the inter-galactic
medium. However, we find that none of the paired galaxies in filaments
and sheets in our sample resides in such massive dark matter halo. At
low masses, the supernova feedback may expel or heat the gas reservoir
and quench the star formation \citep{kaviraj07}. The halo mass may
have a role in shaping the physical properties of the galaxy pairs in
filaments and sheets. But it is difficult to explain the cross-overs
observed in \autoref{Fig3} using these differences in the halo mass
distributions.

 The filaments appear at the intersection of sheets and are generally
 denser compared to the sheets. Studies with N-body simulations
 suggest a successive flow of matter from voids to sheets, sheets to
 filaments and from filaments to clusters \citep{ramachandra,
   galarraga22}. A number of earlier studies find that the galaxy
 pairs are preferentially aligned with the filament axis
 \citep{tempel15, mesa18}. The alignment signal is reported to be
 stronger for closer pairs residing near the filament spine. The
 anisotropic accretion along the filaments may significantly influence
 the gas accretion efficiency in these aligned galaxy pairs and
 trigger interaction induced star formation in them. Contrarily, the
 major pairs with $r>50$ kpc show a lower star formation in filaments
 than in sheets. The filaments are generally denser than the
 sheets. The $D1$-type galaxies are embedded in high density
 environment as compared to the $D2$-type galaxies
 \citep{pandey20}. The galaxies in denser environments are known to be
 redder and low star forming \citep{lewis02, gomez03, kauffmann04}. So
 naively one would expect the galaxies in filamentary environment to
 be less star forming and redder compared to the galaxies in
 sheet-like environment. We find that this is true for the galaxies in
 major pairs with separation larger than $50$ kpc. However, the
 galaxies in major pairs at closer pair separation show a strikingly
 opposite behaviour.

 We do not analyze the alignment of the galaxy pairs in our study. The
 individual sheets and filaments can not be identified using the local
 dimension. We plan to carry out a detailed study of the galaxy pair
 alignment with different identification techniques of the cosmic web
 in a future work.

The Eagle simulation provides two definitions for the position of
galaxies. These are based on the centre of mass and the location of
the minimum of the gravitational potential. The two positions do not
coincide for some galaxies.  In this work, we use the centre of mass
to define the position of galaxies. We also repeat our analysis
considering the minimum of the gravitational potential as the position
of a galaxy. We show the results of this analysis in
\autoref{Fig4}. The main findings of our analysis remain unchanged
with this alternative definition of galaxy position. Further, it is
important to ensure that the major pairs considered in our analysis do
not belong to the galaxy groups. We measure the distances to the
$5^{th}$ nearest neighbours for the paired galaxies in sheets and
filaments and find that $\sim 20 \%$ of them have their $5^{th}$
nearest neighbour within a distance of $500$ kpc to $1$ Mpc. We
discard these galaxy pairs and repeat our analysis. The results of
this analysis are shown in \autoref{Fig5}.  We find that discarding
such galaxy pairs do not alter our results.

The results reported in this work are very similar to the results
obtained in a recent study \citep{das23} of the colour and SFR of
major pairs in filaments and sheets using the SDSS data. \citet{das23}
use volume limited sample of galaxies ($M_r \leq -19$) for their
analysis and find a crossover in these properties at nearly the same
length scale ($\sim 50$ kpc). It is interesting to note that we
observe exactly the same trend in the EAGLE simulation data. This
provides a strong theoretical support to the observational findings
that large-scale structures like sheets and filaments affect galaxy
interactions. This also indicates that the galaxy properties are
modulated by the geometry of their large-scale environment.

 Finally, we conclude that the filaments play a significant role in
 deciding the colour and star formation rate in galaxies. The
 observed differences in the colour and SFR of major pairs in
 filaments and sheets can not be interpreted in terms of the
 differences in the local density and the stellar mass
 distributions. The interacting galaxy pairs with smaller pair
 separation can trigger star formation. The filaments provide a
 favourable environment for such interactions. This makes the
 interacting galaxies bluer in filaments compared to those found in
 sheets.

\begin{acknowledgements}
BP acknowledges financial support from the SERB, DST, Government of
India through the project CRG/2019/001110 and support from IUCAA, Pune
through the associateship programme. SS acknowledges DST, Government
of India for support through a National Post Doctoral Fellowship
(N-PDF).

The authors acknowledge the Virgo Consortium for making their
simulation data publicly available. The EAGLE simulations were
performed using the DiRAC-2 facility at Durham, managed by the ICC,
and the PRACE facility Curie based in France at TGCC, CEA,
Bruy\`{e}res-le-Ch\^{a}tel.

\end{acknowledgements}

\label{lastpage}

\begin{thebibliography}{99}


\bibitem[Alonso et al. (2004)]{alonso04} Alonso, M.~S., Tissera, P.~B., Coldwell, G., Lambas, D.~G., 2004, MNRAS, 352, 1081 




\bibitem[Baldry et al. (2004)]{baldry04} Baldry, I.~K., Glazebrook, K., Brinkmann, J., Ivezi{\'c}, {\v{Z}}., Lupton, R.~H., Nichol, R.~C. \& Szalay, A.~S., ApJ, 600, 681

\bibitem[Balogh, Navarro \& Morris (2000)]{balogh00} Balogh, M.~L., Navarro, J.~F., \& Morris, S.~L., 2000, ApJ, 540, 113 


\bibitem[Bamford, Nichol \& Baldry (2009)]{bamford09} Bamford, S.~P., Nichol, R.~C., Baldry, I.~K., et al., 2009,  MNRAS, 393, 1324


\bibitem[Barton, Geller \& Kenyon (2000)]{barton00} Barton, E.~J., Geller, M.~J., Kenyon, S.~J., 2000, ApJ, 530, 660 

\bibitem[Bhattacharjee, Pandey \& Sarkar (2020)]{bhattacharjee20} Bhattacharjee, S., Pandey, B., \& Sarkar, S., 2020, JCAP, 2020, 039


\bibitem[Birnboim \& Dekel (2003)]{birnboim03} Birnboim Y., Dekel A., 2003, MNRAS, 345, 349


\bibitem[Bond, Kofman \& Pogosyan (1996)]{bond96} Bond, J.~R., Kofman, L., \& Pogosyan, D., 1996, \nat, 380, 603

\bibitem[Bravo-Alfaro et al. (2009)]{bravo09} Bravo-Alfaro H., Caretta C.~A., Lobo C., Durret F., Scott T., 2009, A\&A, 495, 379












\bibitem[Cohen et al. (2014)]{cohen14} Cohen S.~A., Hickox R.~C., Wegner G.~A., Einasto M., Vennik J., 2014, ApJ, 783, 136

\bibitem[Cox et al. (2004)]{cox04} Cox, T.~J., Primack, J., Jonsson, P., \& Somerville, R.~S., 2004, ApJL, 607, L87

\bibitem[\protect\citeauthoryear{Cooray \& Sheth}{2002}]{cooray} Corray, A., Sheth, R.K., \ 2002, Phys. Rep., 371, 1

  
\bibitem[Das, Pandey \& Sarkar (2023)]{das23} Das, A., Pandey, B \& Sarkar, S., 2023, RAA, 02, 23



\bibitem[Galarraga-Espinosa et al. (2021)]{galarraga21} Galarraga-Espinosa, D., Aghanim, N., Langer, M., Tanimura, H., 2021, A\&A, 649, A117 

\bibitem[\protect\citeauthoryear{Gal{\'a}rraga-Espinosa, Garaldi, \& Kauffmann}{2022}]{galarraga22} Gal{\'a}rraga-Espinosa D., Garaldi E., Kauffmann G., 2022, arXiv, arXiv:2209.05495, Accepted in A\&A

\bibitem[Dav{\'e}, Finlator, \& Oppenheimer)(2012)]{dave12} Dav{\'e} R., Finlator K., Oppenheimer B.~D., 2012, MNRAS, 421, 98


\bibitem[Davis \& Geller (1976)]{davis76} Davis, M., \&  Geller, M.J., 1976, ApJ, 208, 13 


\bibitem[Dekel \& Birnboim (2006)]{dekel06} Dekel A., Birnboim Y., 2006, MNRAS, 368, 2


\bibitem[Dekel, Sari, \& Coverino (2009)]{dekel09} Dekel, A., Sari, R., Coverino, D., 2009, ApJ, 703, 785

\bibitem[Doi et al. (2010)]{doi10} Doi, M et al., 2010, doi:10.1088/004-6256/139/4/1628, arxiv:1002.3701


\bibitem[Dressler (1980)]{dress80} Dressler, A., 1980, ApJ, 236, 351






\bibitem[Einasto, Joeveer, \& Saar (1980)]{einasto80} Einasto J., Joeveer M., Saar E., 1980, MNRAS, 193, 353



\bibitem[Ellison et al. (2010)]{ellison10} Ellison, S.~L., Patton, D.~R., Simard, L., McConnachie, A.~W.,Baldry, I.~K., Mendel, J.~T., 2010, MNRAS, 407, 1514


\bibitem[Gabor et al. (2010)]{gabor10} Gabor J.~M., Dav{\'e} R., Finlator K., Oppenheimer B.~D., 2010, MNRAS, 407, 749

\bibitem[Gao et al.(2004)]{gao04} Gao L., White S.~D.~M., Jenkins A., Stoehr F., Springel V., 2004, MNRAS, 355, 819. doi:10.1111/j.1365-2966.2004.08360.x

\bibitem[G{\'o}mez et al. (2003)]{gomez03} G{\'o}mez, P.~L., Nichol, R.~C., Miller, C.~J., Balogh, M.~L., Goto, T., Zabludoff, A.~I., Romer, A.~K., et al., 2003, ApJ, 584, 210

\bibitem[Goto et al. (2003)]{gotto03} Goto, T., Yamauchi, C., Fujita, Y., Okamura, S., Seikiguchi, M., Smail, I., Bernardi, M.,\&  Gomez, P.L., 2003, MNRAS, 346, 601


\bibitem[Gregory \& Thompson (1978)]{gregory78} Gregory S.~A., Thompson L.~A., 1978, ApJ, 222, 784




\bibitem[Gunn \& Gott (1972)]{gunn72} Gunn, J.~E., \& Gott, J.~R., 1972, ApJ, 176, 1 

\bibitem[Guzzo et al. (1997)]{guzo97} Guzzo, L., Strauss, M.A., Fisher,K.B., Giovanelli, R., \&  Haynes, M.P., 1997, ApJ, 489, 37

\bibitem[Helmi (2020)]{helmi20} Helmi A., 2020, ARA\&A, 58, 205



\bibitem[Joeveer \& Einasto (1978)]{joeveer78} Joeveer M., Einasto J., 1978, IAUS, 79, 241


\bibitem[Kaviraj et al. (2007)]{kaviraj07} Kaviraj S., Kirkby L.~A., Silk J., Sarzi M., 2007, MNRAS, 382, 960

\bibitem[Kauffmann et al. (2004)]{kauffmann04}  Kauffmann, G., White, S.~D.~M., Heckman, T.~M., et al., 2004, \mnras, 353, 713

\bibitem[Kauffmann et al. (2003)]{kauffmann03} Kauffmann, G., Heckmann, T, M., White, S, D, M., Charlot, S., Tremonti, C et al., 2003, MNRAS, 341, 54










\bibitem[Larson, Tinsley \& Caldwell (1980)]{larson80} Larson, R.~B., Tinsley, B.~M., \& Caldwell, C.~N., 1980, ApJ, 237, 692





\bibitem[Lewis et al. (2002)]{lewis02}  Lewis, I., Balogh, M., Propris, R. De., Couch, W., Bower, R., Offer, A., Bland-Hawthorn, J., et al., 2002, MNRAS, 334, 673




\bibitem[Martig et al. (2009)]{martig09} Martig M., Bournaud F., Teyssier R., Dekel A., 2009, ApJ, 707, 250


\bibitem[Masters et al. (2010)]{masters10} Masters K.~L., Mosleh M., Romer A.~K., Nichol R.~C., Bamford S.~P., Schawinski K., Lintott C.~J., et al., 2010, MNRAS, 405, 783


\bibitem[McAlpine et al. (2016)]{eagle16} McAlpine, S., Helly, J, C., Schaller, M., Trayford, J, W. et al., 2016, A\&C, 72, 15



\bibitem[Mesa et al. (2018)]{mesa18} Mesa V., Duplancic F., Alonso S., Mu{\~n}oz Jofr{\'e} M.~R., Coldwell G., Lambas D.~G., 2018, A\&A, 619, A24


\bibitem[Moore et al. (1996)]{moore96} Moore, B., Katz, N., Lake, G., Dressler, A., \& Oemler, A., 1996, Nature, 379, 613 

\bibitem[Moore, Lake \& Katz (1998)]{moore98}  Moore, B., Lake, G., \&  Katz, N., 1998, ApJ, 495, 139


\bibitem[Murray, Quataert \& Thompson (2005)]{murray05} Murray, N., Quataert, E., \& Thompson, T.~A., 2005, ApJ, 618, 569

\bibitem[Nikolic, Cullen \& Alexander (2004)]{nikolic04} Nikolic, B., Cullen, H., Alexander, P., 2004, MNRAS, 355, 874 

\bibitem[Oemler(1974)]{oemler74} Oemler, A., 1974, ApJ, 194, 1 

\bibitem[Pandey \& Bharadwaj (2005)]{pandey05} Pandey, B.,  \& Bharadwaj S., 2005, MNRAS, 357, 1068

\bibitem[Pandey \& Bharadwaj (2006)]{pandey06} Pandey, B., \&   Bharadwaj, S., 2006, \mnras, 372, 827 

\bibitem[Pandey \& Bharadwaj (2008)]{pandey08}  Pandey, B., \& Bharadwaj, S., 2008, \mnras, 387, 767 

\bibitem[Pandey et al. (2013)]{pandey13} Pandey B., White S.~D.~M., Springel V., Angulo R.~E., 2013, MNRAS, 435, 2968

\bibitem[Pandey \& Sarkar (2017)]{pandey17} Pandey, B., \&  Sarkar, S., 2017, MNRAS, 467, L6

\bibitem[Pandey \& Sarkar (2020)]{pandey20}  Pandey, B., \& Sarkar, S., 2020, MNRAS, 498, 6069



\bibitem[\protect\citeauthoryear{Park et al.}{2005}]{park1}  Park, C., et al.\ 2005, \apj, 633, 11

\bibitem[Patton et al. (2011)]{patton11} Patton, D.~R., Ellison, S.~L., Simard, L., McConnachie, A.~W., Mendel, J.~T., 2011, MNRAS, 412, 591
 




\bibitem[Peng \& Renzini (2020)]{peng20} Peng Y.-. jie ., Renzini A., 2020, MNRAS, 491, L51


\bibitem[Planck Collaboration et al. (2014)]{planck14} Planck Collaboration et al., 2014,  A\&A, 571, A1 
 



\bibitem[\protect\citeauthoryear{Ramachandra \& Shandarin}{2015}]{ramachandra} Ramachandra N.~S., Shandarin S.~F., 2015, MNRAS, 452, 1643



\bibitem[\protect\citeauthoryear{Saintonge et al.}{2012}]{saintonge12} Saintonge A., Tacconi L.~J., Fabello S., Wang J., Catinella B., Genzel R., Graci{\'a}-Carpio J., et al., 2012, ApJ, 758, 73

\bibitem[Sarkar \& Bharadwaj (2009)]{sarkar09} Sarkar, P, \& Bharadwaj, S., 2009, \mnras, 394, L66

\bibitem[Sarkar \& Pandey (2019)]{sarkar19} Sarkar, S., \& Pandey, B., 2019, MNRAS, 485, 4743 

\bibitem[Sarkar \& Pandey (2020)]{sarkar20}  Sarkar, S., \&  Pandey, B., 2020, MNRAS, 497, 4077





\bibitem[Silk (1977)]{silk77} Silk, J., 1977 ApJ, 211, 638


\bibitem[Somerville \& Primack (1999)]{somerville99} Somerville, R.~S. , \& Primack, J.~R., 1999, MNRAS, 310, 1087 


\bibitem[Springel, Matteo \& Hernquist (2005)]{springel05} Springel, V., Di Matteo, T., \&  Hernquist, L., 2005, MNRAS,361, 776 






  
\bibitem[Tempel \& Tamm (2015)]{tempel15} Tempel E., Tamm A., 2015, A\&A, 576, L5

\bibitem[\protect\citeauthoryear{Thorp et al.}{2022}]{thorp22} Thorp M.~D., Ellison S.~L., Pan H.-A., Lin L., Patton D.~R., Bluck A.~F.~L., Walters D., et al., 2022, MNRAS, 516, 1462

  
\bibitem[Trayford et al. (2015)]{trayford15} Trayford, J, W et al., 2015, MNRAS, 452, 2879

\bibitem[Tuominen (2021)]{tuominen21}  Tuominen T., Nevalainen J., Tempel E., Kuutma T., Wijers N., Schaye J., Hein{\"a}m{\"a}ki P., et al., 2021, A\&A, 646, A156

\bibitem[\protect\citeauthoryear{Violino et al.}{2018}]{violino18} Violino G., Ellison S.~L., Sargent M., Coppin K.~E.~K., Scudder J.~M., Mendel T.~J., Saintonge A., 2018, MNRAS, 476, 2591

\bibitem[White \& Rees (1978)]{white78} White, S.~D.~M., \&  Rees,M.~J., 1978 \mnras, 183, 341 

\bibitem[Woods \& Geller (2007)]{woods07} Woods, D.~F., Geller, M.~J., 2007, AJ, 134, 527 







\bibitem[\protect\citeauthoryear{Zehavi et al.}{2005}]{zehavi} Zehavi, I., et al.\  2005, \apj, 630, 1 

\bibitem[Zeldovich \& Shandarin (1982)]{zeldovich82} Zeldovich I.~B., Shandarin S.~F., 1982, PAZh, 8, 131



\end{thebibliography}
\end{document}